\documentclass[sigconf]{acmart}

\AtBeginDocument{%
\providecommand\BibTeX{{%
Bib\TeX}}}

\AtBeginDocument{%
  \providecommand\BibTeX{{%
    \normalfont B\kern-0.5em{\scshape i\kern-0.25em b}\kern-0.8em\TeX}}}

\copyrightyear{2025}
\acmYear{2025}
\setcopyright{cc}
\setcctype{by}
\acmConference[SIGIR '25]{Proceedings of the 48th International ACM SIGIR Conference on Research and Development in Information Retrieval}{July 13--18, 2025}{Padua, Italy}
\acmBooktitle{Proceedings of the 48th International ACM SIGIR Conference on Research and Development in Information Retrieval (SIGIR '25), July 13--18, 2025, Padua, Italy}\acmDOI{10.1145/3726302.3729879}
\acmISBN{979-8-4007-1592-1/2025/07}

\makeatletter
\gdef\@copyrightpermission{
\begin{minipage}{0.3\columnwidth}
\href{https://creativecommons.org/licenses/by/4.0/}{\includegraphics[width=0.90\textwidth]{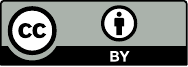}}
\end{minipage}\hfill
\begin{minipage}{0.7\columnwidth}
\href{https://creativecommons.org/licenses/by/4.0/}{This work is licensed under a Creative Commons Attribution International 4.0 License.}
\end{minipage}
\vspace{5pt}
}
\makeatother

\acmSubmissionID{fp0870}

\usepackage{multirow}

\begin{document}

\title{A Generalised and Adaptable Reinforcement Learning Stopping Method}

\author{Reem Bin-Hezam$^{\Diamond,\clubsuit}$}
\email{rybinhezam@pnu.edu.sa}
\orcid{0000-0002-9156-6186}
\affiliation{%
  \institution{$^\clubsuit$Department of Information Systems\\College of Computer and Information Sciences\\Princess Nourah Bint Abdulrahman University}
  \city{Riyadh}
  \country{Saudi Arabia}
}

\author{Mark Stevenson$^\Diamond$}
\email{mark.stevenson@sheffield.ac.uk}
\orcid{0000-0002-9483-6006}
\affiliation{%
  \institution{$^\Diamond$School of Computer Science\\ Faculty of Engineering\\ University of Sheffield}
  \city{Sheffield}
  \country{United Kingdom}}

\def \authors{Reem Bin-Hezam and Mark Stevenson}

\renewcommand{\shortauthors}{Reem Bin-Hezam and Mark Stevenson}

\begin{abstract}

This paper presents a Technology Assisted Review (TAR) stopping approach based on Reinforcement Learning (RL). Previous such approaches offered limited control over stopping behaviour, such as fixing the target recall and tradeoff between preferring to maximise recall or cost. These limitations are overcome by introducing a novel RL environment, GRLStop, that allows a single model to be applied to multiple target recalls, balances the recall/cost tradeoff and integrates a classifier. Experiments were carried out on six benchmark datasets (CLEF e-Health datasets 2017-9, TREC Total Recall, TREC Legal and Reuters RCV1) at multiple target recall levels. Results showed that the proposed approach to be effective compared to multiple baselines in addition to offering greater flexibility.

\end{abstract}



\maketitle

\section{Introduction}

\begin{figure}[!htb] 
  \centering
  \includegraphics[width=0.9999\linewidth]{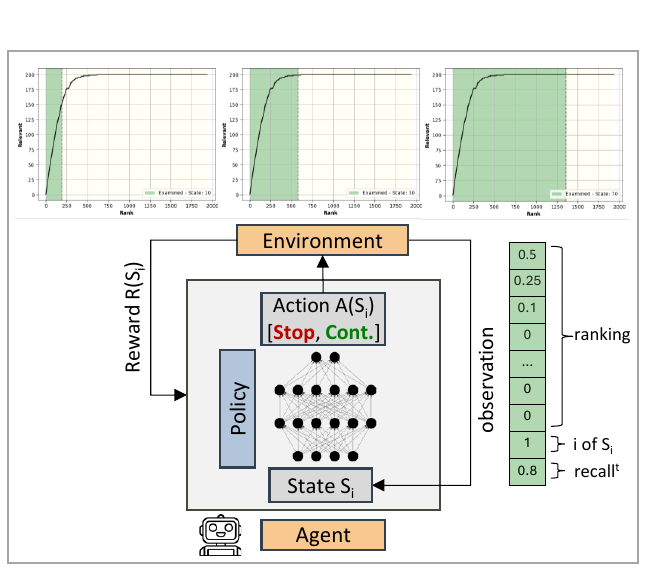} 
  \caption{RL Environment for TAR Stopping} 
  \label{fig:GRLStop_approach_diagram}
\end{figure}

Identifying all, or a significant proportion of, the relevant documents in a collection has applications in multiple areas including development of systematic reviews \cite{higgins2019cochrane,kanoulas2017clef,kanoulas2018clef,kanoulas2019clef}, satisfying legal disclosure requirements \cite{mcdonald2020accuracy,baron2020providing,grossman2016trec}, social media content moderation \cite{yang2021tar} 
and test collection development \cite{Losada2019}. These problems often involve large collections where manually reviewing all documents would be prohibitively time-consuming. Technology Assisted Review (TAR) develops techniques to support these document review processes, including stopping rules which help reviewers to decide when to stop assessing documents, thereby reducing the effort required to screen a collection for relevance. TAR stopping rules aim to identify when a desired level of recall (the {\it target recall}) has been reached during document review, while also minimising the number of documents examined. The problem is challenging since these two objectives are in opposition; increasing the number of documents examined provides more information about whether the target has been reached. 

A wide range of approaches have been applied to the problem, including examination of the rate at which relevant documents are observed \cite{cormack2016engineering,stevenson2023stopping}, estimating the number of remaining relevant documents by sampling or classification \cite{shemilt2014pinpointing,callaghan2020statistical,cormack2016scalability,li2020stop} and analysis of ranking scores \cite{hollmann2017ranking,di2018study}. A recent approach is based on  Reinforcement Learning (RL) (see Figure~\ref{fig:GRLStop_approach_diagram}). {\it RLStop} uses deep RL to train a model to make stopping decisions which were found to perform well in comparison with a wide range of alternative stopping models \cite{binhezam_stevenson_rlstop2024}. However, RLStop suffers from a range of limitations, the most significant being that each model is trained for a specific target recall, limiting their generalisability. Also, in common with many other stopping methods, it does not provide a mechanism to adapt behaviour to balance the two stopping objectives: maximising the likelihood of reaching target recall and minimising the number of documents examined. Finally, RLStop only uses information from documents that have already been examined, despite work demonstrating the value of incorporating predictions about likely relevance of unexamined documents \cite{yu2019fast2,Yang2021heuristic,binhezam_stevenson_emnlp23}.

Inspired by RLStop, this paper proposes an alternative RL-based stopping algorithm which overcomes these limitations and improves performance. In particular, it proposes a new reward function for use within the RL algorithm which allows the creation of a single stopping model that can be used with a range of target recall levels and also provides control over whether proposed stopping points prefer to maximise the number of relevant documents identified or minimise the number of documents examined. In addition, it demonstrates how information from unexamined documents can be incorporated within the RL framework. Experiments on six data sets commonly used in TAR evaluation demonstrate that these extensions improve performance. Further analysis explores the effect of ranking quality.

This work makes the following contributions: (1) introduces a RL-based stopping model that can be applied to multiple target recalls and adapted for different stopping preferences (e.g. maximising reaching target recall or minimising number of documents examined), (2)  integrates a classifier within the RL environment to provide information about the unexamined documents, (3) demonstrates that this approach provides state-of-the-art performance through evaluation using multiple benchmark data sets, and (4) investigates the effect of a ranking quality on these approaches.\footnote{Code available from \url{https://github.com/ReemBinHezam/GRLStop}}

\section{Background}
A wide range of approaches have been proposed for TAR stopping methods. The most common is to estimate the total number of relevant documents in the collection and use this information to determine whether enough documents have been observed for the target recall to have been achieved. The total number of relevant documents has been estimated in a range of ways including sampling the unexamined documents \cite{shemilt2014pinpointing,Howard2020,callaghan2020statistical,cormack2016scalability,li2020stop,molinari2023saltau,bron2024using}, training a classifier using relevance judgments from the examined documents and applying it to those yet to be examined to estimate the number that are relevant \cite{yu2019fast2,yang2021minimizing, Yang2021heuristic}, applying score distribution approaches that make use of the scores assigned by the ranking algorithm \cite{hollmann2017ranking} and by applying counting processes \cite{stevenson2023stopping,binhezam_stevenson_emnlp23}.

Other approaches identify a stopping point without explicitly estimating the total number of relevant documents. The {\it knee method} examines the gain curve produced as relevant documents are encountered to identify an inflection point where these become less frequent \cite{cormack2016engineering}, while {\it target methods} apply sampling theory to randomly sample documents until a pre-specified number of relevant documents have been observed \cite{cormack2016engineering,yang2021minimizing,lewis2021certifying}.

{\it RLStop} also identifies a stopping point without explicitly estimating the total number of relevant documents \cite{binhezam_stevenson_rlstop2024}. RLStop treats the stopping problem as a sequential decision-making problem and employs RL to make repeated decisions to either stop or continue examining documents, in contrast with the more common approach of treating stopping as an estimation problem. Although RL has been widely applied within Information Retrieval \cite{montazeralghaem2020reinforcement,xin2020self,afsar2022reinforcement,ren2023contrastive}, RLStop was the first application to stopping in TAR. RLStop was found to identify suitable stopping points at a range of target recall levels and outperformed other approaches on a range of TAR problems.

Despite its strong performance, RLStop suffers from several limitations. Each model is trained using a single target recall and is intended to make stopping decisions for that recall alone. This limits the generalisability of each model and multiple models would need to be trained in any cases where different target recalls are of interest. This limitation is unusual within stopping methods, for example it is straightforward for approaches based on estimating the total number of relevant documents to adjust the target recall and target methods account of the target recall in the formulae used to determine the number of relevant documents that need to be observed before stopping. 

Applications of TAR stopping rules may have different requirements in terms of the balance between ensuring that the target recall is achieved and minimising the number of documents examined. For example, ensuring that a high proportion of relevant documents are identified is a priority for systematic reviews in the medical domain \cite{higgins2019cochrane}, while stopping as close to the target recall as possible is more likely to be important in the legal domain where minimising cost and unnecessary disclosure are important factors \cite{gray2024high}. Some stopping methods, notably target methods \cite{cormack2016engineering,lewis2021certifying}, do provide this functionality by incorporating the probability that the target recall is reached within the stopping decision. However, RLStop does not allow behaviours to be adapted in this way.  

These limitations are addressed in this work by making use of an alternative reward function within the RL algorithm which allows a single model to be trained and then applied to multiple target recalls. This reward function also allows the balance between ensuring that target recall is achieved and number of documents examined to be considered during training, allowing different models to be created. 

In addition, RLStop examines the documents in the order in which they are ranked and bases its stopping decisions only on information from the documents that have been examined so far. Although this approach has the advantage of prioritising the documents that are more likely to be relevant, information about the unexamined documents has been demonstrated to be useful for this problem \cite{yu2019fast2,yang2021minimizing,binhezam_stevenson_emnlp23}. This paper also demonstrates how information about unexamined documents can be integrated into the RL approach by training a classifier on the documents for which relevance judgments are available and applying it to the unexamined documents.


\section{Reinforcement Learning Approach} 

In RL an agent interacts with an environment and receives rewards depending upon its actions. The environment consists of a {\it state space}, $S$, which defines the set of possible states that the agent can occupy and an {\it action space}, $A$, that lists the set of possible actions that the agent can take at each state. The goal of RL is to learn a {\it policy}, $\pi(s, a)$ where $a \in A$ and $s \in S$, which defines the value of choosing action $a$ given states $s$ and thereby guides the agent's behaviour. A {\it reward function}, $R(s)$, assigns a score to each state and is used to train the policy. An RL algorithm explores the environment by making sequential choices of action through trial-and-error with the goal of maximising the reward obtained. Each choice of action may affect not only the immediate reward obtained but also the rewards obtained following subsequent actions. RL algorithms must balance exploitation and exploration, that is maximising the reward obtained at each step by exploiting what it has already experienced versus maximising cumulative future rewards by exploring the environment by taking new actions \cite{sutton2018reinforcement}.

\subsection{Stopping RL environment}\label{sec:rl_env}

This section describes how a stopping algorithm is implemented within an RL environment. The agent examines a ranked list of documents in batches, starting with the highest ranked documents. After each batch the agent decides to either continue examining documents or stop. 

\noindent{\bf State Space:} The state space represents the document ranking and target recall. A ranking is split into $B$ fixed size batches, each containing $\frac{N}{B}$ documents for a collection of $N$ documents. The agent examines batches sequentially and obtains relevance judgments for all documents in the batch simultaneously. The initial state for each ranking, $S_{1}$, occurs when the first batch (but none of the subsequent batches) has been examined by the agent. Additional batches are examined in subsequent states, i.e. in the $n$th state, $S_{n}$, the first $n$ batches have been examined. The final state, $S_{B}$, represents the situation in which the entire ranking has been examined.

States are represented by a fixed size vector of length $B$+2 where each of the first $B$ elements represent a single batch and the final two represent the number of batches that have been examined so far ($E$) and the target recall.  
The way in which the values representing each batch are computed depends on whether or not the batch has been examined by the agent. For examined batches, i.e. batches $1 \; \ldots \; E$, the value shows the proportion of relevant documents within the batch. However, since the remaining batches, i.e. $E+1 \; \ldots \; N$, have not yet been examined, the number of relevant documents within each is not known. For these batches a classifier is trained using the relevance judgments from the examined batches and applying it to each of the unexamined batches to provide an estimate of the number of relevant documents it contains.

Note that the state representation used by RLStop only included information about relevant documents within examined batches. This work integrates estimates of the number of relevant documents in unexamined batches produced by a classifier, information that has previously been demonstrated to be useful \cite{yu2019fast2,Yang2021heuristic}, into the RL approach.

\noindent{\bf Action Space:} At each point in the ranking, the agent has a choice between two discrete actions: STOP and CONTINUE. The first action is chosen when the agent (informed by the policy) judges that the target recall has been reached. The stopping point returned is the end of the last batch that has been examined so far. If the agent does not stop, it continues to examine the ranking, i.e. moves from state $S_{i}$ to $S_{i+1}$. The last possible agent step is to state $S_{B}$ since this represents the end of the ranking and all documents have been examined.  

\noindent{\bf Policy:} 
A neural network is an appropriate choice of policy when the number of potential states is large, such as the state space encoding used here \cite{Arulkumaran17}. The policy used is a feed-forward network consisting of an input layer of length $B+2$, representing the current state, two 
hidden layers and a binary output layer indicating the chosen action, which is converted to a probability distribution over the two possible actions by a softmax activation function.  

\noindent{\bf Reward function:} The policy is trained using a reward function, $R(S_{i})$, which assigns a score to each state indicating its desirability to the agent. A suitable reward function should have the following properties 1) encourages the agent to continue examining documents until the target recall has been reached, 2) penalises further examination after it has been reached, 3) have the same range for different document rankings and target positions, and 4) can be adapted to offer control over the balance between undershooting and overshooting of target recall. 

Note that the third property is necessary for to develop a single model that can be applied to multiple target recalls since it is difficult for RL algorithms to learn good policies when reward function values across the state space cannot be compared directly.

A function that meets these properties is: 
\begin{equation}\label{eq:stepReward}
R(S_{i}) =  
\begin{cases} 
\frac{i^{m} - (i - 1)^{m}}{T^{m}}   & \text{if } i \leq T \\ \\
\frac{(B - i)^{n} - (B - i + 1)^{n}}{(B - T)^{n}}  & \text{if } i > T
\end{cases} 
\end{equation}

where $i$ is the current state (i.e. $i$th batch), $B$ is the number of batches into which the ranking is split and $T$ is the batch at which the target recall is reached. (Note that while the value of $T$ is known while the algorithm is being trained, it is not known when it is applied.) In addition, $m$ and $n$ are parameters allowing the function to be adapted between preferring to maximise the likelihood of reaching the target recall and minimising the number of documents examined. 

This function assigns a positive reward for states at, or below, the target recall and a negative reward for states after it has been exceeded, thereby meeting the first two properties for a suitable reward function. 

Other properties of Equation~\ref{eq:stepReward} are best understood by considering the {\it cumulative reward} produced for an RL episode, i.e. the sum of rewards for all states visited by the agent during that episode. In this application an episode consists of examining a ranking from the first batch until a stopping decision is made. 
The cumulative reward for an episode that ends at $S_{i}$, $CR(S_{i})$, is given by: 
\begin{equation}\label{eq:cumReward}
 CR(S_{i}) = \begin{cases} 
\left( \frac{i}{T} \right)^{m} & \text{if } i \leq T \\ \\
\left( \frac{B - i}{B - T} \right)^{n} & \text{if } i > T
\end{cases}
\end{equation}

The maximum value for this function, which occurs when the target recall has been reached (i.e. $i = T$) is always 1. The function's value decreases for values below or above $T$ with a minimum possible value of 0 reached when $i = B$ (provided $T \neq B$). The range of the cumulative reward is therefore invariant to the point in the ranking at which the target recall is reached, so the third property is satisfied.

\begin{figure}
\includegraphics[width=0.945\linewidth]{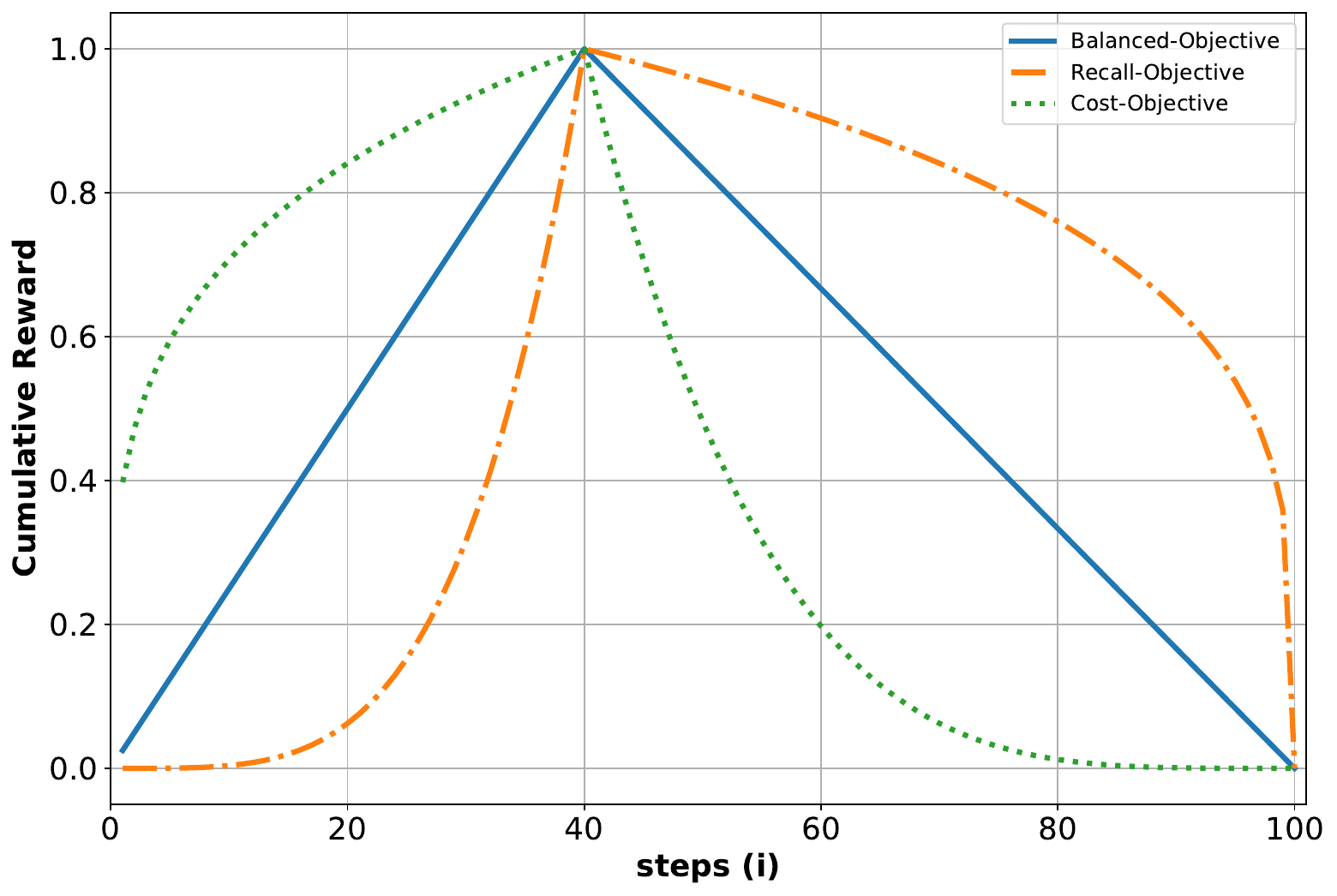}  
  \caption{Example cumulative reward functions produced by varying parameters $m$ and $n$. In this example, the target recall is achieved at batch 40 of 100 in this example (i.e. $T = 40$ and $B = 100$). The solid blue line shows the function produced when $m = n = 1$. The dashed orange line, produced when $m = 4$ and $n = 0.25$, shows a function that assigns less reward before the target recall is achieved and more after it. Similarly, the dotted green line, $m = 0.25$ and $n = 4$, assigns more reward before and less after.} 
  \label{fig:lin_nonlin_rewrd}
\end{figure}

The properties of the reward function, and therefore the cumulative reward, can be controlled by varying the values of the parameters $m$ and $n$. The value of $m$ determines the reward for states before the target recall has been reached. Setting it $< 1$ increases this reward while it is reduced for values $> 1$. Similarly, varying $n$ changes the reward after the target recall has been exceeded.  
Choosing appropriate values for $m$ and $n$ provides a mechanism to control the balance between ensuring the target recall has been reached and minimising the number of documents examined. For example, setting $m = 4$ and $n = 0.25$ alters the function to increase the reward when the target recall is exceeded and reduce the reward when it is not met, thereby encouraging policies that ensure enough documents have been identified to achieve the target recall even at the expense of examining more documents than necessary. Similarly, setting $m < 1$ and $n > 1$ reverses this to increase reward for minimising the number of documents examined, even when the target recall has not been achieved (see Figure \ref{fig:lin_nonlin_rewrd}). 

Note that a reward function could be created using a single parameter to control the balance between maximising recall or minimising workload, and such a function would be adequate for the majority of applications. However, we chose to include two parameters to increase the possible reward functions available. 

The reward function presented here can be compared against the one used by RLStop. That function met the first two properties by assigning positive reward until the target recall was reached and then negative reward thereafter. However, the maximum and minimum cumulative reward varied depending on the batch at which the target recall was reached (i.e. value of $T$) with higher maximum values when this batch occurred later in the rankings. (The maximum cumulative reward for RLStop is $\frac{T -1}{2}$ while the minimum, $\frac{2T -B - 2}{2}$, is negative in some cases.)  
In addition, RLStop's reward function did not offer any mechanism to adapt the policy to prefer to overshoot or undershoot the target recall. 


\section{Experiments} 

\subsection{Datasets}
Performance was evaluated on six datasets from multiple domains that are widely used in high recall retrieval studies. The datasets are highly imbalanced, with a very low percentage of relevant documents for each topic.\\
\noindent{\bf CLEF Technology-Assisted Review in Empirical Medicine (CLEF 2017/2018/2019)} \cite{kanoulas2017clef,kanoulas2018clef,kanoulas2019clef}: A collection of systematic reviews from the Conference and Labs of the Evaluation Forum (CLEF) 2017, 2018, and 2019 e-Health lab Task 2: Technology-Assisted Reviews in Empirical Medicine. The CLEF 2017 dataset contains 42 reviews, CLEF 2018 contains 30 and CLEF 2019 contains 31. The training dataset consists of 12 reviews from CLEF2017.\\
\noindent{\bf TREC Total Recall (TR)} \cite{grossman2016trec}: A collection of 290,099 emails related to Jeb Bush’s eight-year tenure as Governor of Florida (athome4). The collection contains 34 topics. Each topic is labelled with a short title and based on an issue associated with Jeb Bush’s governorship. The training dataset is (athome1) which consists of 10 topics from the same collection.\\
\noindent{\bf TREC Legal (Legal)} \cite{cormack2010overview}: A collection of 685,592 Enron emails made available by the Federal Energy Review Commission during their investigation into the company's collapse. Two topics were used for testing and two for training.\\
\noindent{\bf RCV1}:  \cite{lewis2004rcv1} A collection of Reuters news articles labelled with categories. Following \cite{yang2021minimizing,Yang2021heuristic}, 45 categories were used to represent a range of topics at different prevalence and difficulty levels, and the collections downsampled to 20\% for efficiency. The remaining unused 94 topics in the collection were used for training.

To ensure fair comparison, all stopping methods are applied to the same rankings.
AutoTAR \cite{Cormack2015} is a greedy Active Learning approach that represents state-of-the-art performance on total recall tasks and is commonly used within work on stopping methods. It allows comparison with a range of alternative approaches \cite{li2020stop}. AutoTAR rankings for each dataset were created using a reference implementation and default parameters.\footnote{\url{https://github.com/dli1/auto-stop-tar}} These rankings were used for all experiments, except those reported in Section \ref{sec:rank_quality} which explore the effect of varying ranking quality. 
 

\subsection{Baselines}\label{sec:baselines}

Multiple stopping methods representing a range of  approaches were used as baselines, including those widely used in previous work.

\noindent {\bf RLStop} \cite{binhezam_stevenson_rlstop2024} is the existing RL-based approach described previously. Separate models are trained for each dataset and target recall. \\
{\bf SCAL} \cite{cormack2016scalability} estimates the number of relevant documents in the collection by sampling across the entire ranking.\\
\noindent {\bf AutoStop} \cite{li2020stop} employs a similar approach to SCAL. Sampling is carried out using unbiased estimators \cite{horvitz1952generalization,thompson2012} to account for the decreasing prevalence of relevant documents.\\
\noindent {\bf SD-training/SD-sampling} \cite{hollmann2017ranking} make use of the scores assigned by the ranking algorithm to estimate the total number of relevant documents. They differ in how they identify relevant documents needed to model ranking scores. SD-training from the training data and SD-sampling by sampling documents to obtain relevance judgments from a simulated user.\\
\noindent {\bf IP-H} \cite{stevenson2023stopping} examines the rate at which relevant documents are observed and uses this information within a statistical model (counting process) to estimate the total number of relevant documents.\\
\noindent {\bf Knee} \cite{cormack2016engineering} examines the ``gain curve'' produced by plotting the cumulative total of relevant documents identified against rank. A ``knee detection'' algorithm \cite{satopaa2011finding} is used to examine this curve's gradient to determine when the frequency of relevant documents decreases.\\
\noindent {\bf TM-adapted} \cite{cormack2016engineering} randomly samples from the collection until a pre-specified number of relevant documents have been found. An extension of this approach that allows this figure to be adapted for different target recalls is used \cite{stevenson2023stopping}.\\
\noindent {\bf QBCB} \cite{lewis2021certifying} similar to the previous approach but relies on a set of known relevant documents. Sampling continues until all documents in the set have been found.

Baselines are computed using reference implementations from previous work \cite{li2020stop,stevenson2023stopping} where possible. Otherwise, previously reported results are used and are directly comparable since they are also based on AutoTAR rankings. 
However, some baselines are not available for the RCV1 dataset since we were unable to run the reference code and results have not been provided in previous work. 

Performance was also compared against an {\bf Oracle method (OR)} which examines documents in ranking order and stops when the target recall level has been achieved (or exceeded). The oracle represents the behaviour of an ideal stopping method but is not useful in practise since it requires full information about the ranking. Note that in some cases the oracle can only achieve a recall higher than the target when it is not possible to stop exactly at target, e.g. given a target recall of 0.8 and collection containing 7 relevant documents, the oracle will stop after 6 relevant documents have been found, i.e. recall 0.86.


\subsection{Evaluation Metrics}

Approaches were evaluated using metrics commonly used in previous work on TAR stopping criteria, e.g. \cite{cormack2016engineering,li2020stop}, calculated using  
the {\tt tar\_eval} open-source evaluation script.\footnote{https://github.com/CLEF-TAR/tar}

\noindent {\bf Recall}: Proportion of relevant documents within the collection identified before the method stops examining documents. 

\noindent {\bf Reliability (Rel.)}: Percentage of topics where the desired target recall was reached (or exceeded). For each topic, the reliability is 1 if the target recall is reached before the stopping examining documents, and 0 otherwise. 

\noindent {\bf Cost}: Percentage of documents examined.

\noindent {\bf Cost Difference (CostDiff)}: In addition to the above metrics, the cost difference score is also introduced. This is the difference between the proportion of documents that were examined and would have been examined by the oracle method (i.e. stopping immediately upon reaching the target recall). It is computed as $cost(method) - cost(oracle)$. Positive scores indicate that the target recall has been reached and negative that it was not while the absolute value indicates the proportion of the collection by which the ideal stopping point was missed. Cost difference combines information about whether the target recall has been reached and number of documents examined within a single metric. 


\subsection{Implementation}

Proximal Policy Optimization (PPO) \cite{schulman2017proximal} is used as the RL algorithm. PPO is a policy gradient approach that directly learn the policy that maps states to actions rather than indirectly extracting it from state-action pairs. It is an actor-critic RL algorithm that combines policy-based (actor) and value-based (critic) RL, where the actor network decides the actions, and the critic network evaluates them to optimise the value function (i.e. reward value). PPO is based on the REINFORCE \cite{williams1992simple} algorithm but with several enhancements, and most importantly in our case, collects trajectories from different environments simultaneously from independent parallel actors, which allows a policy to be trained using rankings from multiple topics. PPO employs a clipping mechanism limiting policy updates within a specific range and leading to more stable training which is important when training the agent with multiple environments.

PPO is more suitable than alternative approaches like DQN \cite{mnih2013playing} which relies on Q-values to map actions to specific states and does not easily generalise across different environments such as multiple rankings. PPO is also more sample-efficient than some alternative methods, thereby reducing the amount of data required to learn effective policies.  

\noindent{\bf Classifier:} The classifier applied to the unexamined portion of the ranking was implemented using logistic regression with a TF-IDF document representation. 
Despite its simplicity, this approach has proved successful for TAR problems and has commonly been used by previous approaches \cite{binhezam_stevenson_emnlp23,Yang2021heuristic,li2020stop,yu2019fast2,yang2021minimizing}.
The classifier was implemented using the {\small\texttt{scikit-learn}} 
~library with {\small\texttt{LogisticRegression}} and TF-IDF scores generation using {\small\texttt{TfidfVectorizer}} and default configurations, which proved to work well in previous work \cite{binhezam_stevenson_emnlp23,li2020stop}. Documents features used were title and abstract for the CLEF 2017-19 datasets, title and content of emails for TREC TR and Legal  and the entire news article for RCV1. Cost sensitive learning was used to mitigate class imbalance during classifier training by using a weight of 1 for the minority class (i.e. relevant) and weighing the minority class (i.e. not relevant) as the ratio of the number of documents in the minority and majority classes \cite{ling2008cost}.

\noindent{\bf Implementation and Hyperparameters:}  The RL environment was created using the {\small\texttt{Gymnasium}} library \cite{gymnasium_2023} with vector environments, which allows multiple independent environments to be stacked together, thereby allowing simultaneous training on multiple topics. The {\small\texttt{Stable-Baseline3}} \cite{stable-baselines3} implementation of PPO was used. 

Following previous work, the number of batches, $B$, was set to 100 to ensure a reasonable number of possible stopping positions without the environment becoming too sparse \cite{binhezam_stevenson_rlstop2024}.

Experiments were carried out exploring multiple configurations which required different hyperparameters due to the changes to the environment and reward behaviour. The best values for each setting were identified using a grid search. For all configurations, the number of epochs, discount factor, learning rate, and neural network hidden layers nodes were set to their default values of 10, 0.99, 0.0003 and 64, which also proved to provide the best results. The number of steps per environment at each rollout was set to 10, the entropy coefficient to 0.1 and the clipping range to 0.1. Early stopping callback \cite{stable-baselines3} was applied when there was no improvement in the policy for ten consecutive rollouts to help avoid overfitting, save training time and ensure fair comparison.

Section \ref{sec:classifier} describes a configuration in which the classifier was not included. For this configuration, the number of steps per environment at each rollout was increased to 100 since the environment is less informative and more steps are required for to explore it fully. The entropy coefficient was also set to 0.001 to encourage a balance between the number of steps and exploration, thereby introducing more stable policy updates. 

The reward function parameters $m$ and $n$ were set to 1 for all experiments except those in Section~\ref{sec:objectives} which explore the effect of altering the objectives.

\begin{figure*}[!bt] 
  \centering
  \includegraphics[width=0.999950\linewidth
  ]{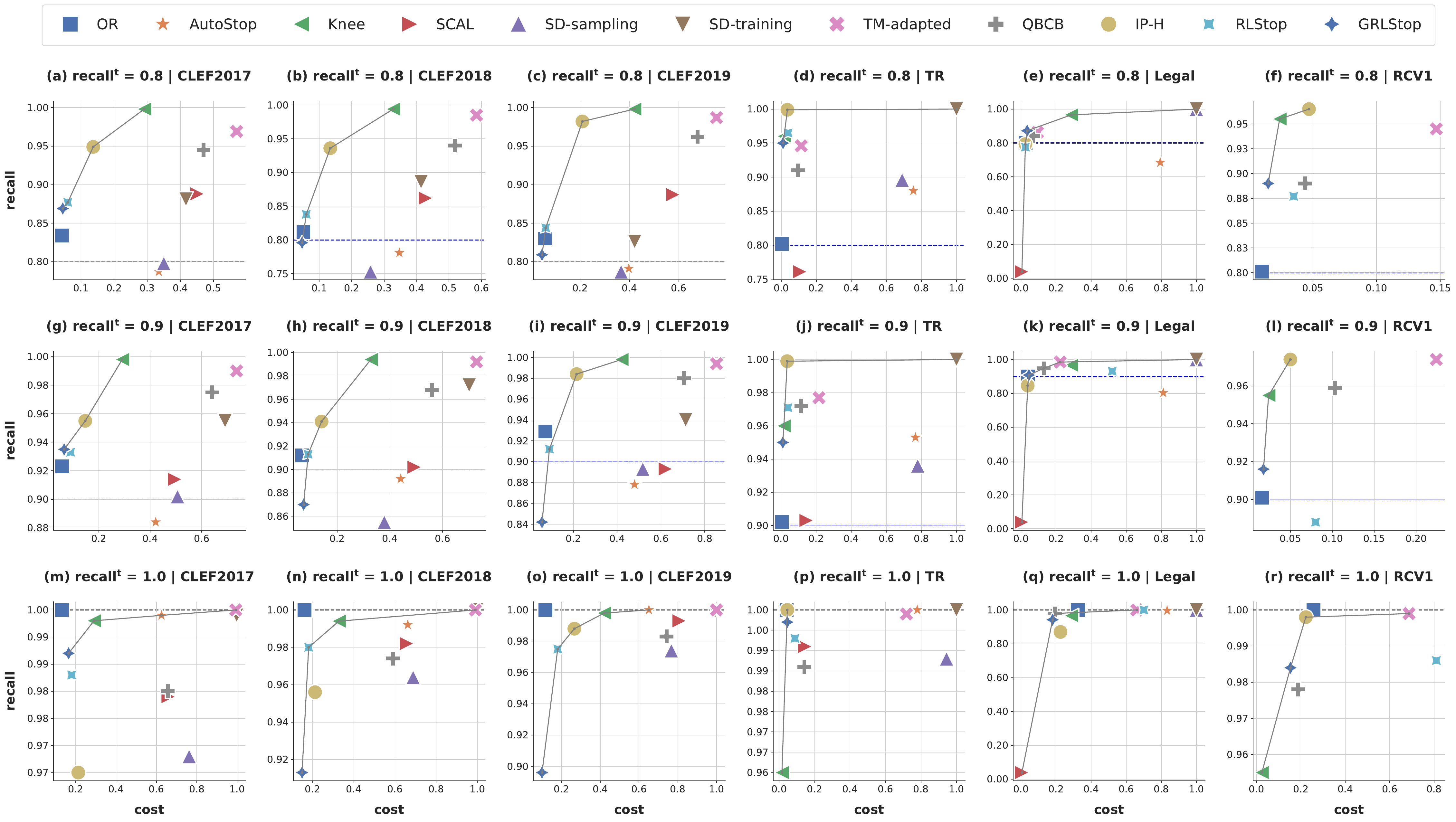}  
  \caption{Recall vs cost. Target recall indicated by horizontal blue dashed line and Pareto front as grey line. Note that scale of y-axis varies across sub-figures.} 
  \label{fig:GRLStop_BLs_grid_recall_cost}
\end{figure*}


\section{Results}

The first experiment compares the proposed approach, referred to as {\bf GRLStop}, with the alternative methods (baselines and oracle) described in Section~\ref{sec:baselines}. Results are reported using recall and cost, averaged across all topics in each collection. (Note that the reliability and CostDiff scores for GRLStop are also available in Table~\ref{tab:generalised_vs_per_target_rlstop}.) Recall and cost metrics were chosen because they provide information about how well the stopping algorithm has achieved its two key objectives: identifying relevant documents and minimising the total number of documents examined. Suitable stopping methods need to take account of both objectives so Pareto efficient approaches (i.e. those that reach the highest recall for a particular cost) are identified for each setting. 

Results are shown in Figure \ref{fig:GRLStop_BLs_grid_recall_cost} for all datasets with target recalls set to 0.8, 0.9 and 1.0. (Target recall 0.7 is also included for subsequent experiments but not for this one since it was not possible to obtain scores for several of the baseline approaches.) Each dataset is represented in a single column and each target recall in a single row. Grey lines in each sub-figure indicate the Pareto front (i.e. set of Pareto efficient approaches). 

The variation in relative performance across the range of configurations included in the experiment demonstrates the difficulty of selecting a single approach that is optimal in all circumstances. However, GRLStop is Pareto optimal in almost every case. The single exception is for the TR dataset for target recall 1.0 (subfigure (p)) where GRLStop is very close to being Pareto optimal. The approach was consistently able to reach the target recall level with lower cost compared to the baselines in almost all non-total recall scenarios when the target recall is 0.8 or 0.9. It is also consistently closer to the optimal Oracle results than other approaches. 

GRLStop often fails to meet the target recall when it is set to 1.0 (bottom row of Figure \ref{fig:GRLStop_BLs_grid_recall_cost}), most noticeably for the CLEF 2018 and 2019 datasets. However, the cost is substantially lower than for other approaches (several of which examine the majority of the collection) and achieves a good balance between cost and recall. 

The knee and IP-H approaches are also Pareto optimal in several cases. However, their cost is generally higher then GRLStop (indicating that they require more documents to be examined) and they tend to be further from the oracle. The remaining approaches are not Pareto optimal under any scenario, or only occasionally.

These results demonstrate that GRLStop is comparable with state-of-the-art approaches for TAR stopping and is often able to achieve performance closer to the optimal oracle than alternative methods.


\subsection{Fixed vs Varying Target Recall}

\begin{table*}[]
\caption{Comparison of GRLStop and RLStop trained on target recalls 0.7 and 0.9}
\label{tab:generalised_vs_per_target_rlstop}
\resizebox{0.9999\textwidth}{!}
{
\begin{tabular}{l|l|lllr|lllr|lllr|lllr}
\toprule
 &  & \multicolumn{4}{|c|}{\textbf{Target Recall = 1.0}} & \multicolumn{4}{|c|}{\textbf{Target Recall = 0.9}} & \multicolumn{4}{|c|}{\textbf{Target Recall = 0.8}} & \multicolumn{4}{|c}{\textbf{Target Recall = 0.7}} \\
\toprule
 \textbf{Dataset} & \textbf{Model} & \textbf{Recall $\uparrow$} & \textbf{Rel. $\uparrow$} & \textbf{Cost $\downarrow$} & \textbf{CostDiff} & \textbf{Recall $\uparrow$} & \textbf{Rel. $\uparrow$} & \textbf{Cost $\downarrow$} & \textbf{CostDiff} & \textbf{Recall} & \textbf{Rel.} & \textbf{Cost} & \textbf{CostDiff} &\textbf{Recall $\uparrow$} & \textbf{Rel. $\uparrow$} & \textbf{Cost $\downarrow$} & \textbf{CostDiff} \\

\midrule
\multirow{3}{*}{\textbf{CLEF2017}} 
 & \textbf{RLStop7} & 0.851 & 0.500 & 0.050 & -0.084 & 0.851 & 0.667 & 0.050 & -0.009 & 0.851 & 0.733 & 0.050 & 0.004 & 0.851 & 0.767 & 0.050 & 0.014 \\
 & \textbf{RLStop9} & 0.933 & 0.600 & 0.090 & -0.044 & 0.933 & 0.767 & 0.090 & 0.031 & 0.933 & 0.800 & 0.090 & 0.044 & 0.933 & 0.933 & 0.090 & 0.054 \\
 & \textbf{GRLStop} & 0.992 & 0.767 & 0.165 & 0.032 & 0.935 & 0.767 & 0.065 & 0.006 & 0.869 & 0.700 & 0.045 & 0.000 & 0.791 & 0.733 & 0.031 & -0.006 \\
\midrule
\multirow{3}{*}{\textbf{CLEF2018}} 
 & \textbf{RLStop7} & 0.803 & 0.333 & 0.050 & -0.110 & 0.803 & 0.500 & 0.050 & -0.019 & 0.803 & 0.667 & 0.050 & -0.004 & 0.803 & 0.700 & 0.050 & 0.006 \\
 & \textbf{RLStop9} & 0.913 & 0.500 & 0.090 & -0.070 & 0.913 & 0.767 & 0.090 & 0.021 & 0.913 & 0.867 & 0.090 & 0.036 & 0.913 & 0.867 & 0.090 & 0.046 \\
 & \textbf{GRLStop} & 0.913 & 0.700 & 0.149 & -0.012 & 0.870 & 0.700 & 0.073 & 0.003 & 0.796 & 0.633 & 0.047 & -0.007 & 0.732 & 0.667 & 0.034 & -0.010 \\
\midrule
\multirow{3}{*}{\textbf{CLEF2019}} 
 & \textbf{RLStop7} & 0.811 & 0.500 & 0.050 & -0.062 & 0.811 & 0.600 & 0.050 & -0.018 & 0.811 & 0.633 & 0.050 & -0.005 & 0.811 & 0.667 & 0.050 & 0.004 \\
 & \textbf{RLStop9} & 0.912 & 0.633 & 0.090 & -0.022 & 0.912 & 0.767 & 0.090 & 0.022 & 0.912 & 0.800 & 0.090 & 0.035 & 0.912 & 0.900 & 0.090 & 0.044 \\
 & \textbf{GRLStop} & 0.896 & 0.633 & 0.099 & -0.013 & 0.842 & 0.700 & 0.056 & -0.012 & 0.809 & 0.667 & 0.045 & -0.010 & 0.743 & 0.633 & 0.034 & -0.011 \\
 \midrule
\multirow{3}{*}{\textbf{Legal}} 
 & \textbf{RLStop7} & 0.504 & 0.000 & 0.010 & -0.320 & 0.504 & 0.000 & 0.010 & -0.035 & 0.504 & 0.000 & 0.010 & -0.025 & 0.504 & 0.000 & 0.010 & -0.015 \\
 & \textbf{RLStop9} & 0.931 & 0.500 & 0.520 & 0.190 & 0.931 & 0.500 & 0.520 & 0.475 & 0.931 & 1.000 & 0.520 & 0.485 & 0.931 & 1.000 & 0.520 & 0.495 \\
 & \textbf{GRLStop} & 0.942 & 0.500 & 0.180 & -0.150 & 0.908 & 0.500 & 0.045 & 0.000 & 0.872 & 1.000 & 0.035 & 0.000 & 0.836 & 1.000 & 0.030 & 0.005 \\
\midrule
\multirow{3}{*}{\textbf{TR}} 
 & \textbf{RLStop7} & 0.965 & 0.324 & 0.039 & -0.010 & 0.965 & 0.971 & 0.039 & 0.027 & 0.965 & 0.971 & 0.039 & 0.028 & 0.965 & 0.971 & 0.039 & 0.028 \\
 & \textbf{RLStop9} & 0.971 & 0.324 & 0.039 & -0.010 & 0.971 & 0.971 & 0.039 & 0.028 & 0.971 & 0.971 & 0.039 & 0.028 & 0.971 & 0.971 & 0.039 & 0.028 \\
 & \textbf{GRLStop} & 0.997 & 0.618 & 0.046 & -0.004 & 0.950 & 0.941 & 0.010 & -0.002 & 0.950 & 0.941 & 0.010 & -0.001 & 0.950 & 0.941 & 0.010 & -0.001 \\
 \midrule
\multirow{3}{*}{\textbf{RCV1}} 
 & \textbf{RLStop7} & 0.850 & 0.156 & 0.034 & -0.229 & 0.850 & 0.533 & 0.034 & 0.012 & 0.850 & 0.689 & 0.034 & 0.017 & 0.850 & 0.800 & 0.034 & 0.020 \\
 & \textbf{RLStop9} & 0.888 & 0.200 & 0.080 & -0.182 & 0.888 & 0.578 & 0.080 & 0.058 & 0.888 & 0.778 & 0.080 & 0.064 & 0.888 & 0.911 & 0.080 & 0.067 \\
 & \textbf{GRLStop} & 0.984 & 0.356 & 0.153 & -0.109 & 0.916 & 0.644 & 0.018 & -0.004 & 0.890 & 0.778 & 0.015 & -0.002 & 0.877 & 0.889 & 0.014 & 0.000 \\
\bottomrule

\end{tabular}
}
\end{table*}

A key feature of GRLStop is its ability to train a single model that can be used to identify stopping points for different target recalls. This is in contrast with RLStop where a different model has to be trained for each target recall, limiting their flexibility. This also provides RLStop with an advantage in the results shown in Figure \ref{fig:GRLStop_BLs_grid_recall_cost} since all GRLStop results for a dataset are produced by a single model, while the RLStop ones are produced by three separate models. 

To explore this further, an experiment was carried out in which GRLStop was compared against versions of RLStop trained for two target recalls: RLStop7 for target recall 0.7 and RLStop9 for target recall 0.9. Each model was then applied in scenarios with for different target recalls (0.7, 0.8, 0.9 and 1.0). The difference between this experiment and the one reported in Figure \ref{fig:GRLStop_BLs_grid_recall_cost} is that the RLStop7 and RLStop9 models are applied to all target recalls, rather than just the one that they have been trained for.

\begin{figure*}[th!] 
  \centering
  \includegraphics[width=0.88\linewidth]{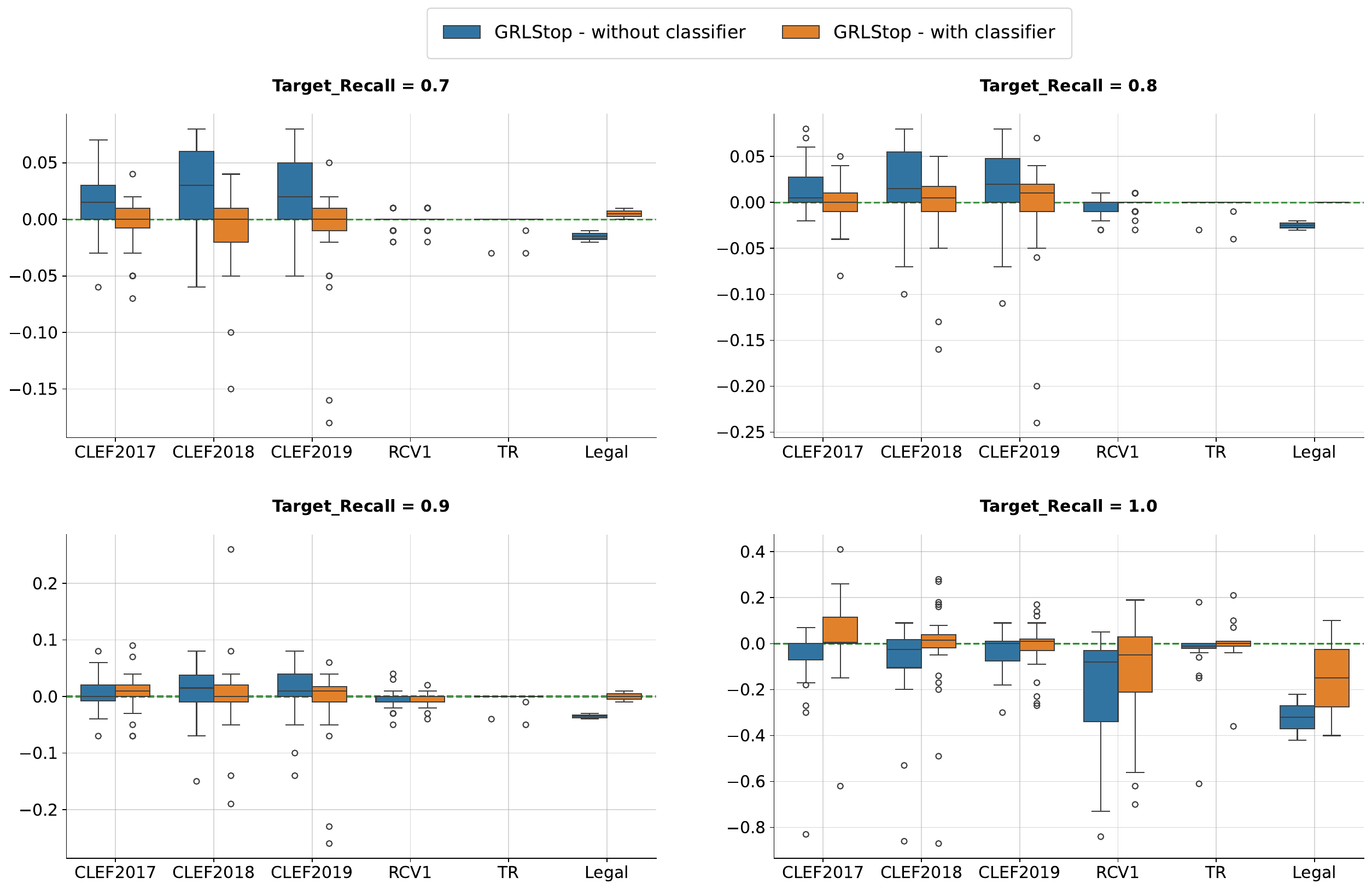}  
  \caption{Effect of including/excluding classifier on CostDiff. Green dashed horizontal line indicates optimal value (i.e. 0).}
  \label{fig:GRLStop_clf_cost}
\end{figure*}

Results are shown in Table \ref{tab:generalised_vs_per_target_rlstop} where recall, reliability, cost and CostDiff metrics are reported for each approach. Note that the recall and cost metrics for the RLStop7 and RLStop9 models are the same for all target recalls, since these models only aim to achieve the target recall they were trained for. The reliability and CostDiff scores for these models do change since these metrics consider the target recall. However, for GRLStop, results for all metrics vary depending on the target recall, demonstrating the models generalisability across target recalls.

Unsurprisingly, the RLStop7 and RLStop9 models perform best when applied to the target recall they had been trained for. But the performance of these models tends to degrade when they are applied to different target recalls. For example, performance of RLStop7 reduces as it is applied to increasingly higher target recalls. In these cases the models tends to undershoot the target, as indicated by the drop in reliability and CostDiff scores. Similarly, RLStop9 overshoots when applied to lower target recalls, as demonstrated by the high cost and cost difference figures in comparison with the other two approaches. 

The costDiff scores for GRLStop are consistently lower than those for RLStop7 and RLStop9, even in the cases where it was more costly (such as target recall 1.0), demonstrating its ability to adapt the stopping decision to the target recall being sought.


\subsection{Classifier Effect}\label{sec:classifier}

A further analysis was carried out to determine the effect of including the classifier to predict relevance of unobserved documents on overall performance. A version of GRLStop that did not employ the classifier was created by adapting the RL environment described by Section \ref{sec:rl_env} so that the classifier-predicted values for each unobserved document (i.e. all batches from $E+1\;\ldots\;N$) are replaced by a dummy value $-1$, an approach similar to the one used by RLStop. 

The box plot in Figure \ref{fig:GRLStop_clf_cost} compares the results obtained with and without the classifier over all topics. CostDiff scores are reported since to allows per-topic differences to be represented. It can be seen that including classifier prediction labels moves the CostDiff score closer to the optimal score of 0 in the majority of scenarios. In the majority of circumstances, particularly for target recalls 0.7 and 0.8 and the CLEF data sets, removing the classifier leads to more documents being examined than necessary, indicated by the overall increase in CostDiff. However, in other cases such as target recall 1.0, removing the classifier results in an increased failure to meet the target recall (as indicated by increased negative CostDiff). 

The differences between the results obtained with and without the classifier for each metric were compared across all target recalls and found to be statistically significant for all target recalls except 0.9 (paired t-test with Bonferroni correction, $p < 0.05$).

These results demonstrate that integrating the classifier's predictions into the RL environment improves stopping decision performance.


\subsection{Adapting Objectives}\label{sec:objectives}

GRLStop allows the reward function to be varied to achieve different objectives such as encouraging the policy to meet the target recall or to minimise the number of documents examined. The effect of doing so was explored in an experiment comparing versions of GRLStop developed using different reward functions. The first, {\it GRLStop-recall-obj}, is designed to encourage a policy that ensures the target recall is achieved, even if this requires more documents to be examined. Its reward function was created by setting $m = 4$ and $n = 0.25$, values chosen to encourage the intended behaviour without being too extreme. The second, {\it GRLStop-cost-obj}, aims to minimise the number of documents examined, even if raises the risk that the target recall is not met. It was created by setting $m = 0.25$ and $n = 4$. These approaches are compared against the standard GRLStop ($m = n = 1$) which balances these objectives ({\it GRLStop-cost-balanced}). The classifier predictions were not included in this experiment since it was found that the impact of varying the reward function was more pronounced when it was not included.

Table \ref{tab:GRLStop-Objectives} shows the performance of the three approaches for a range of target recalls over each data set. These results demonstrate that GRLStop-recall-obj consistently reached higher recall levels than GRLStop and GRLStop-cost-obj, normally at the expense of an increase in cost although this is sometimes quite limited (e.g. for the TR dataset and RCV1 with lower target recalls). The improvement in recall and reliability is particularly noticeable for the Legal dataset, although it only consists of two topics which limits the possible reliability scores. In this case, GRLStop-recall-obj is always able to achieve the target recall for both topics with a cost that does not exceed 20\%. Examination of the learning process revealed that the RL algorithm needed more training steps to converge on a policy for the GRLStop-recall-obj objective, although these policies had higher cumulative rewards than the ones learned using the other two objectives.

On the other hand, the cost for GRLStop-cost-obj is consistently lower than the other two models, often considerably so. This reduction in the number of documents examined comes at the expense of lower recall, which often fails to reach the target. However, the reward function used to train GRLStop-cost-obj is designed to prefer minimising effort over ensuring the target recall has been reached so, in that sense, it has met its objective.

These results demonstrate that adapting the reward function used by GRLStop provides a mechanism to control the balance between preferring to ensure that target recall is achieved and minimising the number of documents examined.

\begin{table*}[!t]
\caption{Effect of Varying Reward Function}
\label{tab:GRLStop-Objectives}
\resizebox{0.9999\textwidth}{!}
{
\begin{tabular}{l|l|lllr|lllr|lllr|lllr}
\toprule
 &  & \multicolumn{4}{|c|}{\textbf{Target Recall = 1.0}} & \multicolumn{4}{|c|}{\textbf{Target Recall = 0.9}} & \multicolumn{4}{|c|}{\textbf{Target Recall = 0.8}} & \multicolumn{4}{|c}{\textbf{Target Recall = 0.7}} \\
\toprule
 \textbf{Dataset} & \textbf{Model} & \textbf{Recall $\uparrow$} & \textbf{Rel. $\uparrow$} & \textbf{Cost $\downarrow$} & \textbf{CostDiff} & \textbf{Recall $\uparrow$} & \textbf{Rel. $\uparrow$} & \textbf{Cost $\downarrow$} & \textbf{CostDiff} & \textbf{Recall $\uparrow$} & \textbf{Rel. $\uparrow$} & \textbf{Cost $\downarrow$} & \textbf{CostDiff} & \textbf{Recall $\uparrow$} & \textbf{Rel. $\uparrow$} & \textbf{Cost $\downarrow$} & \textbf{CostDiff} \\
\midrule
\multirow{3}{*}{\textbf{CLEF2017}} 
 & \textbf{GRLStop-recall-obj} & \textbf{0.967} & 0.700 & 0.129 & -0.005 & \textbf{0.952} & 0.867 & 0.126 & 0.067 & \textbf{0.944} & 0.900 & 0.124 & 0.078 & \textbf{0.937} & 0.967 & 0.116 & 0.080 \\
 & \textbf{GRLStop-balanced} & 0.942 & 0.533 & 0.067 & -0.066 & 0.925 & 0.733 & 0.064 & 0.005 & 0.914 & 0.867 & 0.062 & 0.016 & 0.883 & 0.900 & 0.051 & 0.015 \\
 & \textbf{GRLStop-cost-obj} & 0.685 & 0.267 & \textbf{0.026} & -0.108 & 0.616 & 0.267 & \textbf{0.019} & -0.040 & 0.514 & 0.267 & \textbf{0.010} & -0.036 & 0.514 & 0.367 & \textbf{0.010} & -0.026 \\
\midrule
\multirow{3}{*}{\textbf{CLEF2018}} 
 & \textbf{GRLStop-recall-obj} & \textbf{0.969} & 0.667 & 0.164 & 0.003 & \textbf{0.960} & 0.833 & 0.156 & 0.086 & \textbf{0.942} & 0.900 & 0.140 & 0.086 &\textbf{ 0.941} & 0.967 & 0.136 & 0.092 \\
 & \textbf{GRLStop-balanced} & 0.928 & 0.367 & 0.084 & -0.076 & 0.924 & 0.700 & 0.081 & 0.012 & 0.898 & 0.900 & 0.073 & 0.019 & 0.887 & 0.900 & 0.070 & 0.025 \\
 & \textbf{GRLStop-cost-obj} & 0.686 & 0.167 & \textbf{0.033} & -0.128 & 0.605 & 0.200 & \textbf{0.026} & -0.043 & 0.436 & 0.267 & \textbf{0.010} & -0.044 & 0.436 & 0.300 & \textbf{0.010} & -0.034 \\
\midrule
\multirow{3}{*}{\textbf{CLEF2019}} 
 & \textbf{GRLStop-recall-obj} & \textbf{0.971} & 0.767 & 0.158 & 0.046 &\textbf{ 0.966} & 0.900 & 0.147 & 0.079 & \textbf{0.957} & 0.933 & 0.139 & 0.084 & \textbf{0.952} & 0.967 & 0.136 & 0.090 \\
 & \textbf{GRLStop-balanced} & 0.933 & 0.533 & 0.081 & -0.031 & 0.926 & 0.833 & 0.077 & 0.009 & 0.915 & 0.867 & 0.074 & 0.019 & 0.899 & 0.833 & 0.068 & 0.022 \\
 & \textbf{GRLStop-cost-obj} & 0.661 & 0.367 & \textbf{0.029} & -0.083 & 0.620 & 0.367 & \textbf{0.025} & -0.042 & 0.459 & 0.267 & \textbf{0.010} & -0.045 & 0.459 & 0.267 & \textbf{0.010} & -0.036 \\
\midrule
\multirow{3}{*}{\textbf{Legal}} 
 & \textbf{GRLStop-recall-obj} & \textbf{0.998} & 0.000 & 0.200 & -0.130 & \textbf{0.998} & 1.000 & 0.200 & 0.155 & \textbf{0.998} & 1.000 & 0.200 & 0.165 & \textbf{0.998} & 1.000 & 0.200 & 0.175 \\
 & \textbf{GRLStop-balanced} & 0.504 & 0.000 & \textbf{0.010} & -0.320 & 0.504 & 0.000 & \textbf{0.010} & -0.035 & 0.504 & 0.000 & \textbf{0.010} & -0.025 & 0.504 & 0.000 & \textbf{0.010} & -0.015 \\
 & \textbf{GRLStop-cost-obj} & 0.504 & 0.000 & \textbf{0.010} & -0.320 & 0.504 & 0.000 & \textbf{0.010} & -0.035 & 0.504 & 0.000 & \textbf{0.010} & -0.025 & 0.504 & 0.000 & \textbf{0.010} & -0.015 \\
\midrule
\multirow{3}{*}{\textbf{TR}} 
 & \textbf{GRLStop-recall-obj} & \textbf{0.971} & 0.324 & 0.011 & -0.038 & \textbf{0.970} & 0.971 & 0.011 & -0.001 & \textbf{0.970} & 0.971 & 0.011 & -0.001 & \textbf{0.964} & 0.971 & \textbf{0.010} & -0.001 \\
 & \textbf{GRLStop-balanced} & \textbf{0.971} & 0.324 & 0.016 & -0.033 & \textbf{0.970} & 0.971 & 0.011 & -0.001 & 0.970 & 0.971 & 0.011 & -0.001 & 0.964 & 0.971 & \textbf{0.010} & -0.001 \\
 & \textbf{GRLStop-cost-obj} & 0.950 & 0.294 &\textbf{ 0.010} & -0.039 & 0.950 & 0.941 & \textbf{0.010} & -0.002 & 0.950 & 0.941 & \textbf{0.010} & -0.001 & 0.950 & 0.941 & \textbf{0.010} & -0.001 \\
\midrule
\multirow{3}{*}{\textbf{RCV1}} 
 & \textbf{GRLStop-recall-obj} & \textbf{0.992} & 0.578 & 0.189 & -0.074 & \textbf{0.956} & 0.822 & 0.060 & 0.038 & \textbf{0.922} & 0.844 & 0.023 & 0.007 & \textbf{0.888} & 0.911 & 0.015 & 0.002 \\
 & \textbf{GRLStop-balanced} & 0.976 & 0.222 & 0.066 & -0.197 & 0.898 & 0.578 & 0.017 & -0.005 & 0.872 & 0.733 & 0.013 & -0.003 & 0.850 & 0.800 & 0.012 & -0.002 \\
 & \textbf{GRLStop-cost-obj} & 0.870 & 0.156 & \textbf{0.016} & -0.247 & 0.842 & 0.511 & \textbf{0.013} & -0.010 & 0.802 & 0.556 & \textbf{0.010} & -0.006 & 0.802 & 0.733 & \textbf{0.010} & -0.004 \\
\bottomrule
\end{tabular}
}
\end{table*}


\subsection{Ranking Quality}\label{sec:rank_quality}

\begin{table*}[hbt!]
\caption{Performance on Range of Ranking Qualities}
\label{tab:GRLStop_different_ranking_qualities_wide}
\resizebox{0.995\textwidth}{!}
{
\begin{tabular}{l|l|lllr|lllr|lllr|lllr}
\toprule
 & \textbf{Ranking} & \multicolumn{4}{|c|}{\textbf{Target Recall = 1.0}} & \multicolumn{4}{|c|}{\textbf{Target Recall = 0.9}} & \multicolumn{4}{|c|}{\textbf{Target Recall = 0.8}} & \multicolumn{4}{|c}{\textbf{Target Recall = 0.7}} \\
\toprule
 \textbf{Dataset} & \textbf{Quality} & \textbf{Recall $\uparrow$} & \textbf{Rel. $\uparrow$} & \textbf{Cost $\downarrow$} & \textbf{CostDiff} & \textbf{Recall $\uparrow$} & \textbf{Rel. $\uparrow$} & \textbf{Cost $\downarrow$} & \textbf{CostDiff} & \textbf{Recall $\uparrow$} & \textbf{Rel. $\uparrow$} & \textbf{Cost$\downarrow$} & \textbf{CostDiff} & \textbf{Recall $\uparrow$} & \textbf{Rel. $\uparrow$} & \textbf{Cost $\downarrow$} & \textbf{CostDiff} \\
 
\midrule
\multirow{3}{*}{\textbf{CLEF2017}} 
 & \textbf{\textbf{Low}}   & 0.941 & 0.567 & 0.423 & 0.005 & 0.898 & 0.667 & 0.235 & 0.079 & 0.856 & 0.733 & 0.133 & 0.051 & 0.767 & 0.767 & 0.061 & 0.005 \\
 & \textbf{\textbf{Mid}}   & 0.961 & 0.700 & 0.228 & 0.002 & 0.935 & {\bf 0.833} & 0.071 & 0.015 & 0.885 & {\bf 0.800} & {\bf 0.048} & 0.005 & {\bf 0.837} & {\bf 0.833} & 0.037 & 0.002 \\
 & \textbf{\textbf{Good}}  & {\bf 0.989} & {\bf 0.767} & {\bf 0.226} & 0.066 & {\bf 0.957} & {\bf 0.833} & {\bf 0.067} & 0.008 & {\bf 0.890} & 0.767 & {\bf 0.048} & 0.002 & 0.809 & 0.767 & {\bf 0.035} & -0.002 \\
\midrule
\multirow{3}{*}{\textbf{CLEF2018}} 
 & \textbf{\textbf{Low}}  & {\bf 0.988} & 0.633 & 0.498 & 0.078 & {\bf 0.942} & {\bf 0.833} & 0.259 & 0.110 & {\bf 0.897} & {\bf 0.900} & 0.141 & 0.057 & 0.805 & 0.800 & 0.091 & 0.029 \\
 & \textbf{\textbf{Mid}}  & 0.973 & 0.700 & 0.235 & -0.011 & 0.892 & 0.700 & 0.073 & -0.007 & 0.842 & 0.800 & 0.054 & 0.002 & {\bf 0.822} & {\bf 0.833} & 0.046 & 0.002 \\
 & \textbf{\textbf{Good}}  & 0.954 & {\bf 0.733} & {\bf 0.191} & -0.018 & 0.876 & 0.667 & {\bf 0.066} & -0.002 & 0.832 & 0.733 & {\bf 0.050} & -0.005 & 0.734 & 0.600 & {\bf 0.034} & -0.011 \\
\midrule
\multirow{3}{*}{\textbf{CLEF2019}} 
 & \textbf{\textbf{Low}}  & {\bf 0.911} & {\bf 0.600} & 0.493 & 0.081 & 0.848 & 0.633 & 0.321 & 0.083 & 0.798 &  0.633 & 0.175 & 0.017 & 0.732 & 0.633 & 0.120 & 0.002 \\
 & \textbf{\textbf{Mid}}  & 0.866 & 0.567 & 0.192 & -0.080 & 0.842 & 0.700 & 0.077 & -0.024 & {\bf 0.802} & {\bf 0.667} & 0.058 & -0.015 & {\bf 0.811} & {\bf 0.800} & 0.054 & -0.012 \\
 & \textbf{\textbf{Good}}  & 0.882 & {\bf 0.600} & {\bf 0.168} & -0.074 & {\bf 0.854} & {\bf 0.733} & {\bf 0.067} & -0.022 & 0.789 & {\bf 0.667} & {\bf 0.048} & -0.028 & 0.750 & 0.667 & {\bf 0.037} & -0.031 \\
\bottomrule
\end{tabular}
}
\end{table*}

Results for the majority of stopping methods have only been reported for a single ranking, despite previous work demonstrating the ranking quality can affect stopping method performance \cite{stevenson2023stopping}. GRLStop's performance across ranking qualities was explored by generating three sets of rankings with different levels of effectiveness. These were produced using AutoTAR with low, mid-range and good rankings being created by stopping the active learning process after assessing 25\%, 50\% and 75\%  of the collection respectively. Quality of the rankings produced was measured by computing the area under recall curve (AURC) and found to average 0.87, 0.92 and 0.96 for the low, mid and good rankings. For comparison, the equivalent score for the rankings used in the previous experiments was 0.97. 

Table \ref{tab:GRLStop_different_ranking_qualities_wide} shows the results of GRLStop on the CLEF datasets. Results for other data sets display similar trends but are not included for brevity. 
These results indicate that, as expected, GRLStop's performance is affected by the ranking quality. The most noticeable difference is in the cost scores which increase as the ranking quality declines, with a particular jump between mid and low quality rankings. For example, for the CLEF2017 data set with target recall 1.0 the cost increases from 0.226 to 0.228 when moving from high to mid quality rankings but then to 0.423 for low quality. The reason for this increase is that the relevant documents occur later in poorer rankings, forcing GRLStop to progress further down the ranking before stopping. 

GRLStop displays some robustness to ranking quality since this does not appear to affect reliability scores which, although they vary with ranking quality, do not always decrease for lower quality rankings (e.g. CLEF2018 dataset for target recalls 0.9 and 0.8). However, there is a general trend for the CostDiff scores to increase as ranking quality decreases, indicating that the approach is forced to examine more documents than necessary before stopping. This is probably because lower quality rankings represent a more noisy environment and therefore a more challenging one for the RL algorithm to learn a good policy for.


\section{Conclusion}

This paper introduces a generalised and adaptable RL-based stopping method for TAR stopping approach. Unlike previous RL-based approaches, the proposed approach allows a single model to be applied to multiple target recall levels and the balance between minimising cost versus achieving target recall to be controlled. It also demonstrates how the output from a text classifier can be used within an RL-based stopping framework.

Results on several benchmark datasets showed that the proposed approach proved to be effective compared to multiple baselines including a previous RL-based method. Further experiments demonstrated that the approach achieved its goals of being applicable to multiple target recall levels and controlling objectives. The integration of text classification was also found to be beneficial.

Possibilities for future work include making use of curriculum learning to reduce the time required to train RL models \cite{bengio2009curriculum} and experimenting with LLMs to produce relevance judgments for the unexamined batches as an alternative to training a text classifier.

%
\bibliographystyle{splncs04}
\bibliography{GRLStop_references}

\end{document}